\documentclass{pasj00}
\draft

\begin{document}
\SetRunningHead{J. W. Lee et al.}{Physical Properties of TrES-3}

\title{Physical Properties of the Transiting Planetary System TrES-3}

\author{Jae Woo \textsc{Lee}, Jae-Hyuck \textsc{Youn}, Seung-Lee \textsc{Kim}, Chung-Uk \textsc{Lee}, and Jae-Rim \textsc{Koo}}
\affil{Korea Astronomy and Space Science Institute, Daejeon, 305-348, Korea}
\email{jwlee@kasi.re.kr, jhyoon$@$kasi.re.kr, slkim@kasi.re.kr, leecu@kasi.re.kr, koojr@kasi.re.kr}


\maketitle

\begin{abstract}
We present four new transits of the planetary system TrES-3 observed between 2009 May and 2010 June. Among these, 
the third transit by itself indicates possible evidence for brightness disturbance, which might be the result of 
the planet blocking a cool starspot on the stellar surface. A total of 109 transit times, including our measurements, 
were used to determine the improved ephemeris with a transit epoch of 2454185.910944$\pm$0.000072 HJED 
and an orbital period of 1.30618700$\pm$0.00000015 d. We analyzed the transit light curves using the JKTEBOP code 
and adopting the quadratic limb-darkening law. In order to derive the physical properties of the TrES-3 system, 
the transit parameters are combined with the empirical relations from eclipsing binary stars and 
stellar evolutionary models. The stellar mass and radius obtained from a calibration using $T_A$, log $\rho_{\rm A}$ 
and [Fe/H] are consistent with those from the isochrone analysis. We found that the exoplanet TrES-3b has a mass 
of 1.93$\pm$0.07 M$_{\rm Jup}$, a radius of 1.30$\pm$0.04 R$_{\rm Jup}$, a surface gravity of 
log $g_{\rm b}$=3.45$\pm$0.02, a density of 0.82$\pm$0.06 $\rho_{\rm Jup}$, and an equilibrium temperature of 
1641$\pm$23 K. The results are in good agreement with theoretical models for gas giant planets.
\end{abstract}

\section{Introduction}

Since the first discovery of a transiting planet around HD 209485 (Charbonneau et al.\ 2000; Henry et al.\ 2000), 
about 100 extrasolar planets have been found to transit their parent stars during the past decade. 
Transit light curves provide us the opportunity to measure the relative size of both the star and planet, 
the orbital inclination, and the stellar limb-darkening coefficients. Together with radial-velocity measurements, 
the observations of transits allow the precise stellar and planetary parameters to be determined and then provide 
constraints on fundamental models of planet formation and evolution. This is similar to the situations 
earlier found in well-detached eclipsing and single-lined spectroscopic binaries. Planetary transits have been 
mainly modeled via the analytical formulae given by Mandel \& Agol (2002) and Gim\'enez (2006), both of which 
require the approximations that the planets are spherical. Because planetary systems are a special case of 
binary star systems, their transits can be analyzed by eclipsing binary models based on numerical integration. 
Gim\'enez (2006) and Southworth (2008) have shown that the detached binary code using biaxial ellipsoidals works 
very well for the study of transiting planets.

The present paper is concerned with the extrasolar planetary system TrES-3 consisting of a nearby G-type dwarf and 
a massive hot Jupiter. This was discovered to be a transiting planet with a short orbital period of about 31 hours, 
a mass of 1.90 M$_{\rm Jup}$, and a radius of 1.30 R$_{\rm Jup}$ by the Trans-atlantic Exoplanet Survey 
(TrES) network (O'Donovan et al.\ 2007). Since then, the system has been the subject of several investigations and 
its observational history was reviewed in a recent paper by Southworth (2010). In this work, we report and analyze 
four new transits of the TrES-3 system and present improved system parameters based on all data collected 
from the literature and obtained in our CCD photometry. One of our main intentions is to show that the work done 
in the field of eclipsing binaries can be applied satisfactorily to transiting planetary systems.

\section{Observations}

We observed four transits of TrES-3 at two observing sites, Sobaeksan Optical Astronomy Observatory (SOAO) in Korea 
and the Mt. Lemmon Optical Astronomy Observatory (LOAO) in Arizona, USA. The first transit was observed on the night 
of 2009 May 7 using a SITe 2K CCD camera attached to the 61-cm reflector at SOAO. The other three were observed
between 2010 April and June using an FLI IMG4301E CCD camera attached to the 1.0 m reflector at LOAO. A summary of 
the observations is given in Table 1, where we present observing interval, site, filter, binning mode, exposure time, 
and numbers of observed points. The instrument and reduction method for each site have been described by 
Lee et al.\ (2007) and Lee et al.\ (2009b) in the same order.

In order to make an artificial comparison source that would be optimal for each transit, we monitored a few tens 
of stars imaged on the chip at the same time as the observing target. After potentially useful field stars were 
examined in detail for any peculiar light variations, about 10 candidate comparisons were combined by 
a weighted average.  Then, the transit light curves from the artificial reference star were normalized by 
fitting a linear function of time to the out-of-transit data to remove time-varying atmospheric effects. 
Four resultant transits are displayed in Figure 1 as differential magnitudes {\it versus} HJD.

\section{Light-Curve Analysis and Transit Times}

As shown in Figure 1, the observed duration and depth of the LOAO $I$ transit is shorter and shallower than those 
of the $R$ band by about 6 min and 0.003 mag, respectively. For the same $R$ band, the third transit is longer 
in duration and deeper in depth than the fourth one and also displays a possible brightening around 
the mid-transit time. If this feature is real, it would be ascribed to a large starspot or a complex of smaller spots 
occulted by the planet (Silva 2003; Rabus et al. 2009) and thus would be evidence for magnetic activity on the surface 
of the parent star. As with the observed variations in eclipse times of binary stars with convective outer layers 
(cf.\ Lee et al.\ 2009b), such activity may cause transit timings to be shifted from conjunction instants.

To derive the transit parameters of TrES-3, we modeled its light curves using the JKTEBOP code 
(Southworth et al.\ 2004a,b), where the SOAO data were used only to obtain a mid-transit time because of large scatter.
The code is based on the EBOP program (Etzel 1981; Popper \& Etzel 1981) originally developed to simulate 
detached eclipsing binary stars using biaxial spheroids. A transit light curve is dependent on the orbital period ($P$), 
the fractional radius of the star ($r_{\rm A}$=$R_{\rm A}$/$a$), the ratio of the radii ($k$=$r_{\rm b}/r_{\rm A}$),
the orbital inclination ($i$), and the limb-darkening coefficients (LDCs) to the host star, where $R_{\rm A}$ is 
the stellar radius and $a$ is the orbital semi-major axis. Thoughout this paper, we refer to the star and planet of 
the TrES-3 system with the subscripts `A' and `b', respectively. 
 
In this analysis, we assumed a circular orbit (Fressin et al.\ 2010; Croll et al.\ 2010) and used $P$=1.30618700 d 
determined from our analysis of transit timings below. Initial quadratic LDCs were taken from the tables of 
Claret (2000), using the atmospheric parameters of $T_{\rm A}$=5650$\pm$75 K, log $g_{\rm A}$=4.4$\pm$0.1, and 
[Fe/H]=$-$0.19$\pm$0.08 (Sozzetti et al.\ 2009). We used $r_{\rm A}$+$r_{\rm b}$ and $k$ as the fitting parameters, 
because these parameters are more weakly correlated than between $r_{\rm A}$ and $r_{\rm b}$ and between $r_{\rm A}$ 
and $r_{\rm A}$+$r_{\rm b}$ (Southworth 2008). The best results for each bandpass are given in columns (2) and (3) 
of Table 2, where the $R$-band solution was obtained from fitting the linear LDC ($u_A$) but fixing the non-linear LDC 
($v_A$). Figure 2 displayed the light curves with best-fitting models and residuals. We chose the weighted means of 
individual solutions listed in the fourth column of the table as our final transit parameters. From those values 
and the stellar velocity amplitude ($K_A$=369$\pm$11 m s$^{-1}$) of Sozzetti et al.\ (2009), we computed 
the stellar density of $\rho_{\rm A}$=1.639$\pm$0.050 $\rho_\odot$ and the planetary surface gravity and zero-albedo 
equilibrium temperature of $g_{\rm b}$=28.2$\pm$1.1 m s$^{-2}$ and $T_{\rm eq}$=1641$\pm$23 K, respectively.

We derived the mid-transit time for each light curve with the JKTEBOP code and the transit parameters of Table 2. 
The results are given in Table 3 together with those taken from Sozzetti et al.\ (2009), Gibson et al.\ (2009), and 
the Exoplanet Transit Database\footnote {http://var2.astro.cz/ETD/} (Poddan\'y et al.\ 2010), where the times are 
linked to be the coordinated universal timescale (UTC). As advocated by Bastian (2000), we transformed 
the HJD timings to the Terrestrial Time (TT) scale so as to use a uniform time system. These are listed as HJED 
(Heliocentric Julian Ephemeris Date) in the second column of the table. The entire collection of 109 transit timings 
from the literature and our observations has been used to determine the following refined ephemeris of TrES-3: 
\begin{equation}
 C_{\rm tr} = \mbox{HJED}~ 2,454,185.910944(72) + 1.30618700(15)E, 
\end{equation}
where $E$ is the number of orbital cycles elapsed from the reference epoch and the parenthesized numbers are 
the 1$\sigma$-error values for the last digit of each term of the ephemeris. The uncertainties of individual times
were taken as weights. The linear least-squares fit yields $\chi^2$=264.98 for 107 degrees of freedom. Our period 
is somewhat longer than those of Sozzetti et al.\ (2009, $P$=1.3061858$\pm$0.0000005 d) and 
Gibson et al.\ (2009, $P$=1.3061864$\pm$0.0000005 d).

The observed ($O$) {\it minus} calculated ($C_{\rm tr}$) mid-transit times from this ephemeris are given in 
the fifth column of Table 3 and plotted in Figure 3, wherein our timings are marked by the filled circles. 
As can be seen in the figure, these residuals show a short-term scatter of about $\pm$0.0025 d, which is several times 
larger than the typical timing precision ($<0.001$ d). This could be produced by the presence of additional planets 
in the TrES-3 system (e.g., WASP-3, Maciejewski et al.\ 2010), which is similar to the case of two planets 
around the binary star system HW Vir discovered by using eclipse timings (Lee et al.\ 2009a). 
In order to see if the residuals represent real and periodic variations, we used the discrete Fourier transform 
program PERIOD04 (Lenz \& Breger 2005). The power spectra indicate a frequency of 0.507 c/d (corresponding to 1.97 d) 
close to the outer 3:2 mean motion resonance but it is difficult to confirm this periodicity because of 
the very low-amplitude signal to noise ratio of 2.8. It is possible that the timing scatter may arise from the presence 
of surface inhomogeneities linked to magnetic activity such as a starspot.

\section{Results and Discussion}

Double-lined spectroscopic and eclipsing binaries can provide a direct determination of the mass and radius 
for each component star. However, because transiting planetary systems display in general radial velocities 
of only the parent stars, their absolute dimensions cannot be determined directly from the observed quantities. 
In order to obtain the physical properties of the TrES-3 system, we have applied both the parameters calculated 
above and the spectroscopic measurements ($K_A$, $T_{\rm A}$, [Fe/H]) of Sozzetti et al.\ (2009) 
to the empirical relations from eclipsing binary stars and stellar evolutionary models. The planet velocity 
amplitude $K_{\rm b}$ was used as a variable governing the process to find the best match between the observations 
and predictions, as in the method described by Southworth (2009). We performed $K_{\rm b}$ searches over 
the range of 50 to 250 km s$^{-1}$ in our subsequent analyses.

First of all, we calculated the physical properties of the transiting system using the mass--radius and 
mass--temperature (hereafter M--R--T) relations from well-studied eclipsing binaries (Southworth 2009). 
The procedure is to look for the stellar mass and hence $K_{\rm b}$ satisfying simultaneously the two relations. 
This consists of calculating the $\chi^2$ fitting statistic,
\begin{equation}
\chi^2 = \biggl[ {{r_{\rm A} - (R_{\rm A,pred} / a)} \over {\sigma_{r_{\rm A}}}} \biggr]^2
+ \biggl[ {{T_{\rm A} - T_{\rm A,pred}} \over {\sigma_{T_{\rm A}}}} \biggr]^2,
\end{equation}
where $\sigma_{r_{\rm A}}$ and $\sigma_{T_{\rm A}}$ are the uncertainties corresponding to the observed values
of $r_{\rm A}$ and $T_{\rm A}$, and $R_{\rm A,pred}$ and $T_{\rm A,pred}$ are the relation-predicted values.
The $K_{\rm b}$-search results appear as the dotted curve in Figure 4 and the system parameters are listed 
in the second column of Table 4. Here, the quantity $\Theta$ denotes the Safronov (1972) number.

A recent study by Torres et al.\ (2010) showed that accurate values for the stellar masses and radii 
could be estimated from $T_{\rm A}$, log $g_{\rm A}$, and [Fe/H]. Enoch et al.\ (2010, hereafter ECPH) 
suggested a new calibration, replacing log $g_{\rm A}$ with log $\rho_{\rm A}$, and showed that the results 
of the relations are consistent with those known for the host stars of 
17 WASP\footnote {Wide Angle Search for Planets, http://www.superwasp.org/} transiting planets. 
In transiting systems, stellar densities are obtained from the analysis of light curves directly and precisely 
(Seager \& Mall\'en-Ornelas 2003). Accordingly, the calibrated mass and radius of 
$M_{\rm A,ECPH}$= 0.968$\pm$0.017 M$_\odot$ and $R_{\rm A,ECPH}$= 0.822$\pm$0.011 R$_\odot$ were used 
to minimize the $\chi^2$ expressed as:
\begin{equation}
\chi^2 = \biggl[ {{M_{\rm A,ECPH} - M_{\rm A,pred}} \over {\sigma_{M_{\rm A,ECPH}}}} \biggr]^2
+ \biggl[ {{R_{\rm A,ECPH} - R_{\rm A,pred}} \over {\sigma_{R_{\rm A,ECPH}}}} \biggr]^2,
\end{equation}
where $M_{\rm A,pred}$ and $R_{\rm A,pred}$ are the stellar mass and radius predicted from the observed quantities 
and the $K_{\rm b}$ values appearing as the dashed curve in Figure 4. The results are given in the third column 
of Table 4.

At the final stage, we used the stellar evolutionary models from the Yonsei--Yale (Y$^2$) series (Yi et al.\ 2001; 
Demarque et al.\ 2004) to constrain the physical properties of the TrES-3 system. A series of the isochrones 
was extracted by considering both the metallicities allowed by the observational errors in [Fe/H] and the ages 
in step of 0.1 Gyr for each metallicity. The provisional mass $M_{\rm A}$ from $K_{\rm b}$ was interpolated 
within the Y$^2$ models to obtain the model-predicted radius $R_{\rm A,pred}$ and temperature $T_{\rm A,pred}$. 
This process also finds the velocity amplitude of $K_{\rm b}$ such that the observed properties of the star 
are best fitted to the predictions of theoretical models by re-using Equation (2) over the range of $K_{\rm b}$ 
from 142 to 250 km s$^{-1}$, because stellar models are not available for masses below 0.40 M$_\odot$. 
Our final results are displayed as the solid curve in Figure 4 and listed in the fourth column of Table 4. 
Despite its large error, the evolutionary age indicates that the transiting system is likely to be young.

We determined the stellar and planetary parameters of the TrES-3 system using the planet velocity amplitude 
$K_{\rm b}$ as a key parameter, following the basic concept and procedure of Southworth (2009). As listed in 
Table 4, the physical properties based on the Y$^2$ evolutionary models are consistent with the results 
using the ECPH calibration within the uncertainties, but not with those of the M--R--T relation. This indicates 
that the empirical calibration on $T_A$, log $\rho_{\rm A}$ and [Fe/H] allows one to infer the masses and radii 
of extrasolar parent stars from the observed values and is a valid alternative to stellar isochrone analysis.
New and independent estimates for the system parameters confirm the previously published properties of this system. 

Our results from the Y$^2$ isochrones indicate that TrES-3A hosts a transiting planet with a mass of 
1.93 M$_{\rm Jup}$, a radius of 1.30 R$_{\rm Jup}$, a semi-major axis of 0.023 AU, and an age of 0.2 Gyr. 
We compared the mass and radius to the theoretical predictions with an orbital separation of 0.02 AU for 300 Myr 
given by Fortney et al.\ (2007), whose models offer the predicted radii for planets of various masses, core masses, 
orbital distances, and ages, orbiting a solar-like star. The planetary radii interpolated from their tabulated values 
are 1.32, 1.30, 1.28, 1.24, and 1.16 R$_{\rm Jup}$ for the models with no core, and core masses of 10, 25, 50, 
and 100 M$_{\oplus}$, respectively. Thus, our measured radius is within the expected radii for gas giant planets 
and TrES-3b has a core mass less than $\sim$38 M$_{\oplus}$. This agrees with the prediction of Burrows et al.\ (2007) 
that there is a positive correlation between a planetary core mass and its host star's metallicity. 

By combining our four timings with data from the literature, we obtained an improved transit ephemeris, which will 
be used as a reference for future study. More accurate and continuous transit observations will help to identify 
and understand the possible transiting timing variations of this planetary system.

\bigskip

The authors wish to thank the staffs of SOAO and LOAO for assistance with our observations. This research has made use 
of the Simbad database maintained at CDS, Strasbourg, France.

\clearpage
\begin{figure}
  \begin{center}
    \FigureFile(150mm,150mm){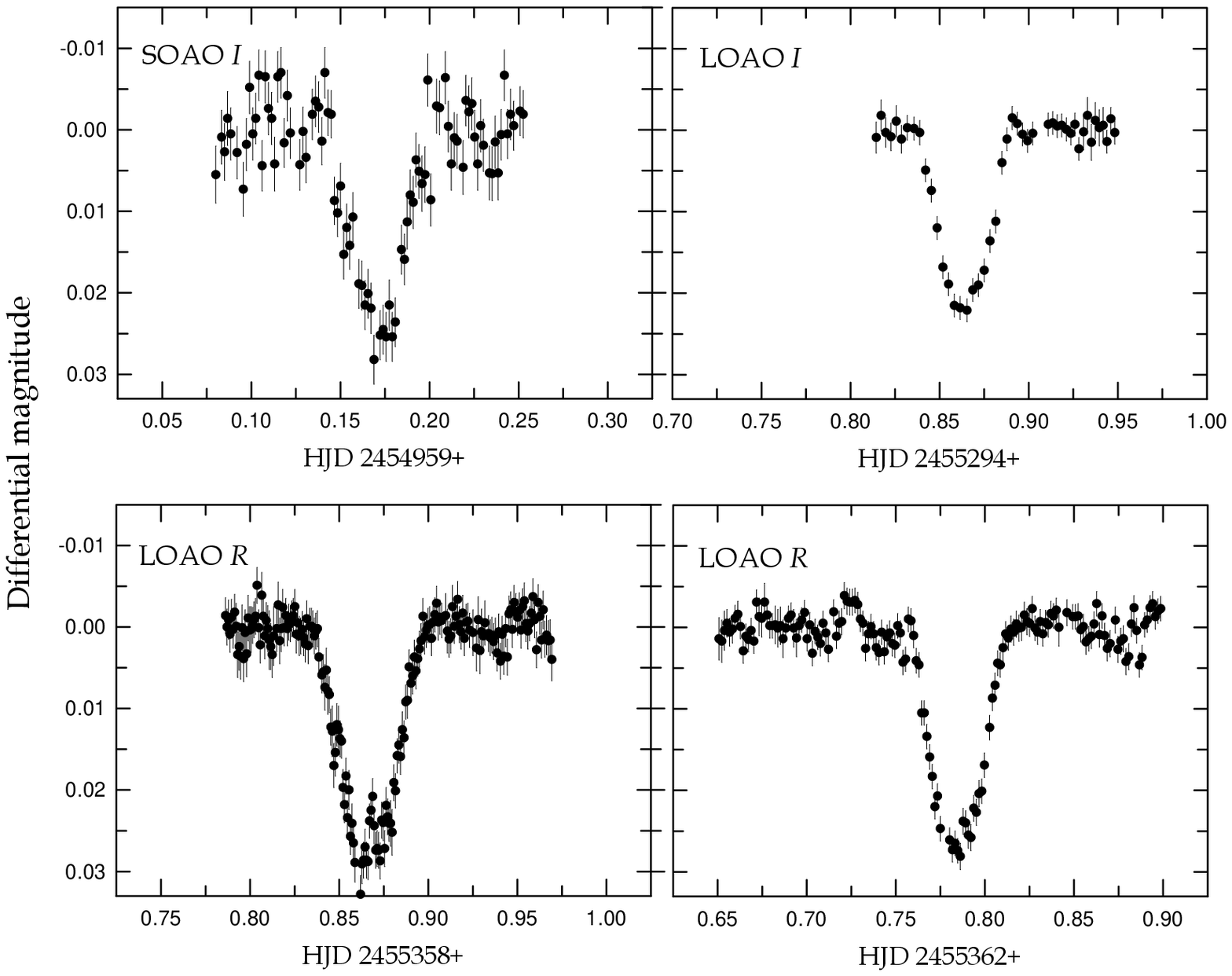}
  \end{center}
  \caption{Light curves of TrES-3 in individual runs. Each transit is labeled by the telescope and filter used.}
  \label{Fig1}
\end{figure}

\begin{figure}
  \begin{center}
    \FigureFile(150mm,150mm){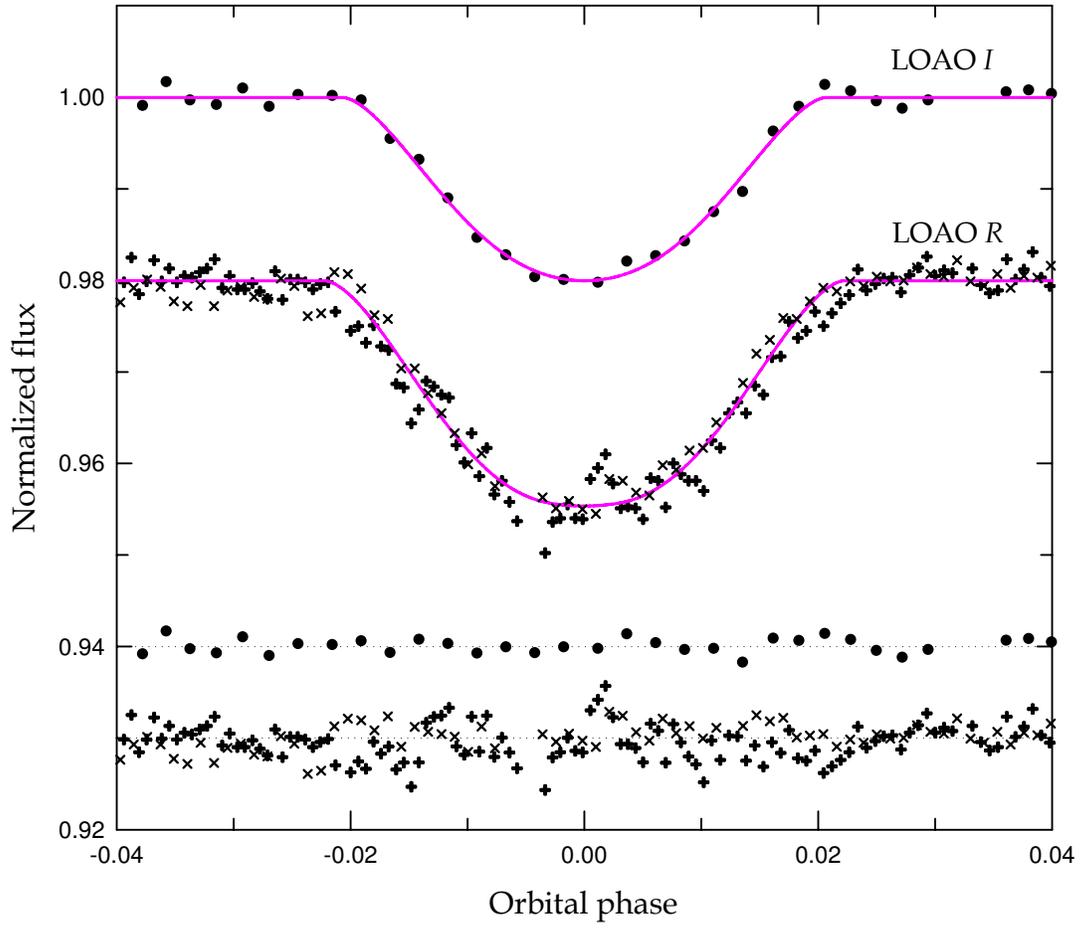}
  \end{center}
  \caption{Phased light curves of TrES-3. LOAO $R$ curve is displaced vertically for clarity. The continuous curves 
  represent the solutions obtained with the best-fit parameters listed in Table 2. The residuals from the fit are 
  offset from zero and plotted at the bottom in the same order as the transit curves. The circle, plus, and `x' symbols 
  indicate the second to fourth transits, respectively.}
  \label{Fig2}
\end{figure}

\begin{figure}
  \begin{center}
    \FigureFile(150mm,150mm){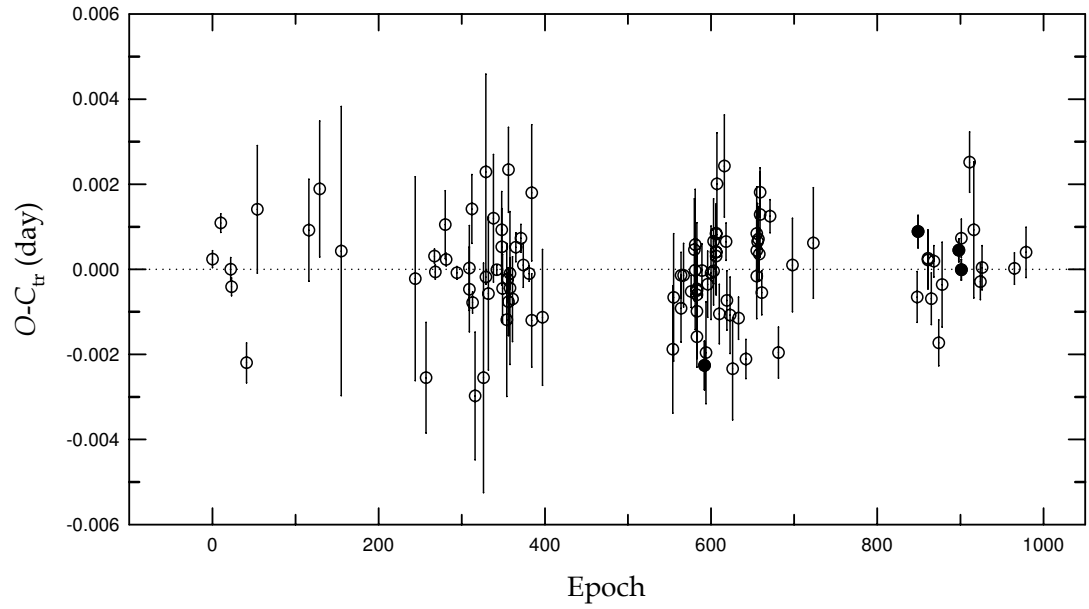}
  \end{center}
  \caption{$O-C_{\rm tr}$ residuals for TrES-3b constructed with the transit ephemeris newly derived in this work. 
  Literature data are plotted as the open circles while our measurements as the filled ones.}
  \label{Fig3}
\end{figure}

\begin{figure}
  \begin{center}
    \FigureFile(150mm,150mm){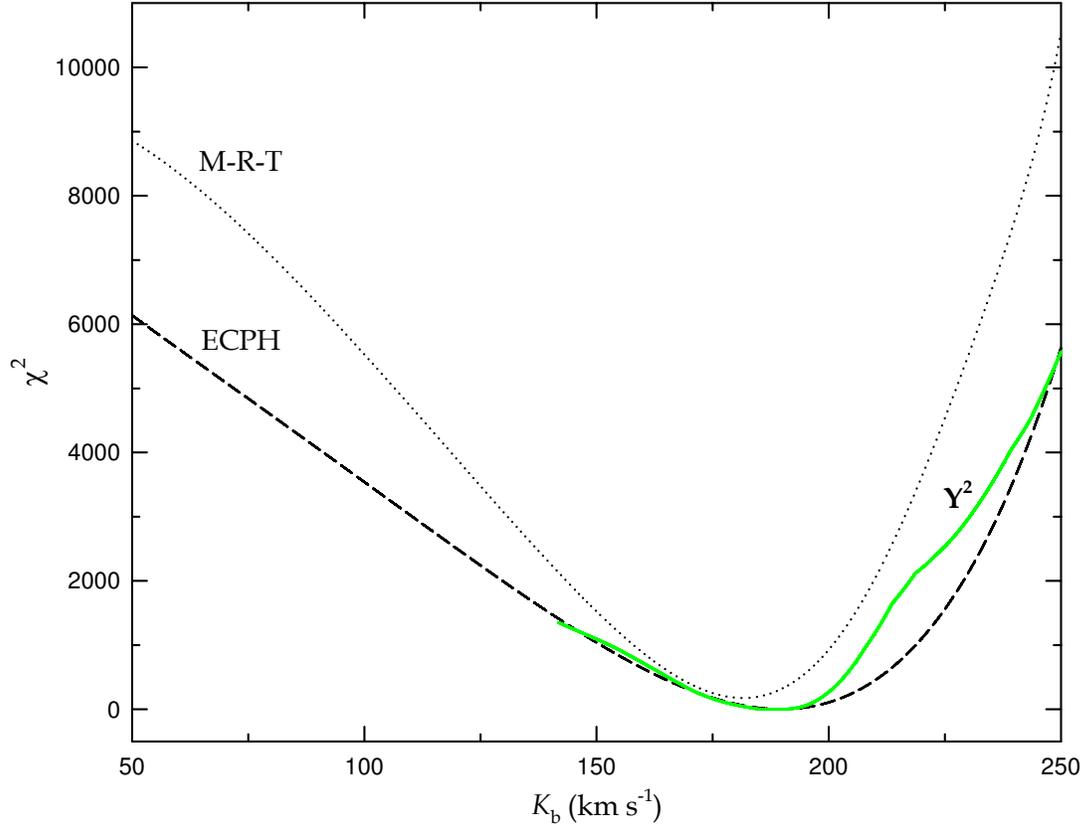}
  \end{center}
  \caption{Behavior of $\chi^2$ as a function of $K_{\rm b}$.  The dotted, dashed, and solid curves represent 
  the $K_{\rm b}$-search results obtained from the M--R--T relation, the ECPH calibration, and the Y$^2$ model, 
  respectively, showing a minimum value at 181.6, 189.8, and 188.8 km s$^{-1}$ in the same order.}
  \label{Fig4}
\end{figure}

\clearpage
\begin{table}
\caption{Observing log of TrES-3.}
\label{tab1}
\begin{center}
\begin{tabular}{llcccccc}
\hline\hline
Transit            &  UT Date            &  Observing Interval            &  Site            &  Filter            &  Binning            &  Exposure            &  $N_{\rm obs}$      \\
                   &                     &  (HJD+2,450,000)               &                  &                    &  Mode               &  Time (s)            &                     \\ \hline
1                  &  2009 05 07         &  4,959.07$-$4,959.25           &  SOAO            &  $I$               &  2$\times$2         &  90                  &  88                 \\  
2                  &  2010 04 08         &  5,294.81$-$5,294.95           &  LOAO            &  $I$               &  1$\times$1         &  160-250             &  46                 \\  
3                  &  2010 06 11         &  5,358.78$-$5,358.97           &  LOAO            &  $R$               &  2$\times$2         &  60-80               &  196                \\  
4                  &  2010 06 15         &  5,362.77$-$5,362.90           &  LOAO            &  $R$               &  1$\times$1         &  100                 &  158                \\ \hline
\end{tabular}
\end{center}
\end{table}

\clearpage
\begin{table}
\caption{Transit parameters of TrES-3.}
\label{tab2}
\begin{center}
\begin{tabular}{lccc}
\hline\hline
Parameter                     &  LOAO $I$                 &  LOAO $R$                   &  Weighted Mean             \\ \hline
$r_{\rm A} + r_{\rm b}$       &  0.1957$\pm$0.0021        &  0.1950$\pm$0.0025          &  0.1954$\pm$0.0023         \\
$k$ (=$r_{\rm b}/r_{\rm A}$)  &  0.1556$\pm$0.0042        &  0.1650$\pm$0.0042          &  0.1603$\pm$0.0042         \\
$i$ (deg)                     &  81.47$\pm$0.14           &  82.06$\pm$0.14             &  81.77$\pm$0.14            \\
$u_A$                         &  0.304                    &  0.456$\pm$0.069            &                            \\
$v_A$                         &  0.300                    &  0.264                      &                            \\
$T_0$ (HJD)$^{\dagger}$       &  5294.86383$\pm$0.00038   &  5358.86636$\pm$0.00018     &                            \\
$r_{\rm A}$                   &  0.1693$\pm$0.0015        &  0.1674$\pm$0.0023          &  0.1687$\pm$0.0017         \\
$r_{\rm b}$                   &  0.02635$\pm$0.00084      &  0.02763$\pm$0.00068        &  0.02712$\pm$0.00074       \\ \hline
\multicolumn{4}{l}{$\dagger$ HJD 2,450,000 is suppressed.} \\
\end{tabular}
\end{center}
\end{table}

\clearpage
\begin{table}
\caption{Observed transit times of TrES-3. A sample is shown here: the full version is available in its entirety in a machine-readable form in the online journal.}
\label{tab3}
\begin{center}
\begin{tabular}{llcrrl}
\hline\hline
HJD          &  HJED$\rm^{a}$ &  Uncertainty   &  $E$  &  $O$--$C_{\rm tr}$ & References$\rm^{b}$            \\
(2,450,000+) &  (2,450,000+)  &                &       &                    &                                \\ \hline
4,185.91043  &  4,185.91118   &  $\pm$0.00020  &    0  &     0.00024        &  Sozzetti et al. (2009)        \\ 
4,198.97315  &  4,198.97390   &  $\pm$0.00022  &   10  &     0.00109        &  Sozzetti et al. (2009)        \\
4,214.64630  &  4,214.64705   &  $\pm$0.00028  &   22  &     0.00000        &  Sozzetti et al. (2009)        \\
4,215.95208  &  4,215.95283   &  $\pm$0.00021  &   23  &  $-$0.00041        &  Sozzetti et al. (2009)        \\
4,239.46166  &  4,239.46241   &  $\pm$0.00047  &   41  &  $-$0.00220        &  Fabjan (TRESCA)               \\
4,256.4457   &  4,256.44645   &  $\pm$0.00150  &   54  &     0.00141        &  Vanmunster (AXA)              \\
4,337.4288   &  4,337.42955   &  $\pm$0.00120  &  116  &     0.00092        &  Hentunen (AXA)                \\
4,354.4102   &  4,354.41095   &  $\pm$0.00160  &  129  &     0.00189        &  Bel (AXA)                     \\
4,388.3696   &  4,388.37035   &  $\pm$0.00340  &  155  &     0.00043        &  Dufoer (AXA)                  \\
4,504.6196   &  4,504.62035   &  $\pm$0.00240  &  244  &  $-$0.00022        &  Naves (AXA)                   \\
4,521.5977   &  4,521.59845   &  $\pm$0.00130  &  257  &  $-$0.00255        &  Br\'at (TRESCA)               \\ \hline
\multicolumn{6}{l}{$^a$ HJD in the terrestrial time scale.} \\
\multicolumn{6}{l}{$^b$ AXA (Amateur eXoplanet Archieve), TRESCA (TRansiting ExoplanetS and CAndidates).} \\
\end{tabular}
\end{center}
\end{table}

\clearpage
\begin{table}
\caption{Physical properties of the TrES-3 system.}
\label{tab4}
\begin{center}
\begin{tabular}{lccc}
\hline\hline
Parameter                        &  M--R--T                    & ECPH                        & Y$^2$ Model                 \\ \hline                                                       
$K_{\rm b}$ (km s$^{-1}$)        &  181.6$\pm$8.2              &  189.8$\pm$3.0              &  188.8$\pm$2.3              \\
$M_{\rm A}$ (M$_\odot$)          &  0.839$\pm$0.085            &  0.958$\pm$0.034            &  0.944$\pm$0.025            \\
$R_{\rm A}$ (R$_\odot$)          &  0.800$\pm$0.037            &  0.836$\pm$0.016            &  0.832$\pm$0.013            \\
log $g_{\rm A}$ (cgs)            &  4.555$\pm$0.059            &  4.574$\pm$0.022            &  4.572$\pm$0.018            \\
$\rho_{\rm A} (\rho_\odot)$      &  1.639$\pm$0.050            &  1.639$\pm$0.050            &  1.639$\pm$0.050            \\
$M_{\rm b}$ (M$_{\rm Jup}$)      &  1.786$\pm$0.170            &  1.950$\pm$0.085            &  1.931$\pm$0.074            \\
$R_{\rm b}$ (R$_{\rm Jup}$)      &  1.252$\pm$0.066            &  1.308$\pm$0.041            &  1.302$\pm$0.038            \\
log $g_{\rm b}$ (cgs)            &  3.451$\pm$0.018            &  3.451$\pm$0.018            &  3.451$\pm$0.018            \\
$\rho_{\rm b} (\rho_{\rm Jup})$  &  0.850$\pm$0.111            &  0.813$\pm$0.060            &  0.817$\pm$0.056            \\
$T_{\rm eq}$ (K)                 &  1641$\pm$23                &  1641$\pm$23                &  1641$\pm$23                \\
$\Theta$                         &  0.0749$\pm$0.0106          &  0.0717$\pm$0.0045          &  0.0721$\pm$0.0039          \\
$a$ (AU)                         &  0.02209$\pm$0.00099        &  0.02308$\pm$0.00037        &  0.02297$\pm$0.00028        \\
Age (Gyr)                        &                             &                             &  0.2$\pm$1.0                \\ \hline
\end{tabular}
\end{center}
\end{table}


\clearpage
\begin{thebibliography}{}
\bibitem[Bastian(2000)]{bastian2000} Bastian, U. 2000, Inf. Bull. Variable Stars, No. 4822
\bibitem[Burrows et al(2007)]{burrows2007} Burrows, A., Hubeny, I., Budaj, J., \& Hubbard, W. B. 2007, ApJ, 661, 502
\bibitem[Charbonneau et al(2000)]{charbonneau2000} Charbonneau, D., Brown, T. M., Latham, D. W., \& Mayor, M. 2000, ApJ, 529, L45
\bibitem[Claret(2000)]{claret2000} Claret, A. 2000, A\&A, 363, 1081
\bibitem[Croll(2010)]{croll2010} Croll, B., Jayawardhana, R., Fortney, J., Lafreniere, D., \& Albert, L. 2010, ApJ, 718, 920
\bibitem[Demarque et al(2004)]{demarque2004} Demarque, P., Woo, J.-H., Kim, Y.-C., \& Yi, S. K. 2004, ApJS, 155, 667
\bibitem[Enoch et al(2010)]{enoch2010} Enoch, B., Cameron, A. C., Parley, N. R., \& Hebb, L. H. 2010, A\&A, 516, A33 (ECPH)
\bibitem[Etzel(1981)]{etzel1981} Etzel, P. B. 1981, in Photometric and Spectroscopic Binary Systems, eds. E. B. Carling and Z. Kopal (Dordrecht: Reidel), 111
\bibitem[Fortney et al(2007)]{fortney2007} Fortney, J. J., Marley, M. S., \& Barnes, J. W. 2007, ApJ, 659, 1661
\bibitem[Fressin et al(2010)]{fressin2010} Fressin, F., et al. 2010, ApJ, 711, 374
\bibitem[Gibson et al(2009)]{gibson2009} Gibson, N. P., et al. 2009, ApJ, 700, 1078
\bibitem[Gim\'enez(2006)]{gimenez2006} Gim\'enez, A., 2006, A\&A, 450, 1231
\bibitem[Henry et al(2000)]{henry2000} Henry, G. W., Marcy, G. W., Butler, R. P., \& Vogt, S. S. 2000, ApJ, 529, L41
\bibitem[Lee et al(2007)]{lee2007} Lee, J. W., Kim, C.-H., \& Koch, R. H. 2007, MNRAS, 379, 1665
\bibitem[Lee et al(2009a)]{lee2009a} Lee, J. W., Kim S.-L., Kim C.-H., Koch R. H., Lee C.-U., Kim H.-I., \& Park J.-H. 2009a, AJ, 137, 3181
\bibitem[Lee et al(2009b)]{lee2009b} Lee, J. W., Youn, J.-H., Lee, C.-U., Kim, S.-L., \& Koch, R. H. 2009b, AJ, 138, 478
\bibitem[Lenz \& Breger(2005)]{lenz2005} Lenz, P., \& Breger, M. 2005, Comm. Asteroseismology, 146, 53
\bibitem[Maciejewski et al(2010)]{2010aciejewski} Maciejewski, G., et al. 2010, MNRAS, 407, 2625
\bibitem[Mandel \& Agol(2002)]{mandel2002} Mandel, K., \& Agol, E. 2002, ApJ, 580, L171
\bibitem[O'Donovan et al(2007)]{odonovan2007} O'Donovan, F. T., et al. 2007, ApJ, 663, L37
\bibitem[Poddany et al(2010)]{poddany2010} Poddan\'y, S., Br\'at, L., \& Pejcha, O. 2010, New Astron., 15, 297
\bibitem[Popper \& Etzel(1981)]{popper1981} Popper, D. M., \& Etzel, P. B. 1981, AJ, 86, 102
\bibitem[Rabus et al(2009)]{rabus2009} Rabus, M., et al. 2009, A\&A, 494, 391
\bibitem[Safronov(1972)]{safronov1972} Safronov, V. S. 1972, Evolution of the Protoplanetary Cloud and Formation of the Earth and Planets (Jerusalem: Israel Program for Scientific Translation)
\bibitem[Seager \& Mallen-Ornelas(2003)]{seager2003} Seager, S., \& Mall\'en-Ornelas, G. 2003, ApJ, 585, 1038
\bibitem[Silva(2003)]{silva2003} Silva, A. V. R. 2003, ApJ, 585, L147
\bibitem[Southworth(2008)]{southworth2008} Southworth, J. 2008, MNRAS, 386, 1644
\bibitem[Southworth(2009)]{southworth2009} Southworth, J. 2009, MNRAS, 394, 272
\bibitem[Southworth(2010)]{southworth2010} Southworth, J. 2010, MNRAS, 408, 1689
\bibitem[Southworth et al(2004a)]{southworth2004a} Southworth, J., Maxted, P. F. L., \& Smalley, B. 2004a, MNRAS, 349, 547
\bibitem[Southworth et al(2004b)]{southworth2004b} Southworth, J., Maxted, P. F. L., \& Smalley, B. 2004b, MNRAS, 351, 1277
\bibitem[Sozzetti et al(2009)]{sozzetti2009} Sozzetti, A., et al. 2009, ApJ, 691, 1145
\bibitem[Torres et al(2010)]{torres2010} Torres, G., Andersen, J., \& Gim\'enez, A. 2010, A\&ARv, 18, 67
\bibitem[Yi et al(2001)]{yi2001} Yi, S., Demarque, P., Kim, Y.-C., Lee, Y.-W., Ree, C. H., Lejeune, T., \& Barnes, S. 2001, ApJS, 136, 417
\end{thebibliography}
\end{document}